# Minimum Spanning Trees in Weakly Dynamic Graphs

Moustafa Nakechbandi
Jean-Yves Colin
Le Havre University, LITIS
France
moustafa.nakechbandi@univ-lehavre.fr
jean-yves.colin@univ-lehavre.fr

Hervé Mathieu
Le Havre University, ISEL
France
herve.mathieu@univ-lehavre.fr

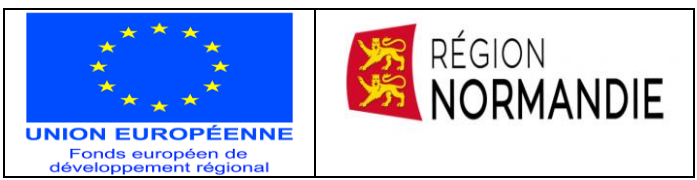

*Abstract*— **In this paper, we study weakly dynamic undirected graphs, that can be used to represent some logistic networks. The goal is to deliver all the delivery points in the network. The network exists in a mostly stable environment, except for a few edges known to be non-stable. The weight of each of these non-stable edges may change at any time (bascule or lift bridge, elevator, traffic congestion...). All other edges have stable weights that never change. This problem can be now considered as a Minimum Spanning Tree (MST) problem on a dynamic graph. We propose an efficient polynomial algorithm that computes in advance alternative MSTs for all possible configurations. No additional computation is then needed after any change in the problem because the MSTs are already known in all cases. We use these results to compute critical values for the non-stable weights and to pre-compute best paths. When the non-stable weights change, the appropriate MST may then directly and immediately be used without any recomputation.**

## I. Introduction

Static graphs have a long history of being used to efficiently represent static problems. In these problems, all the data are known from the start. The real world is not static, however, and the solutions to static problems may not always be used [13], [2]. Some data may change, or be unknown in advance. In territorial systems, for example, the traversal duration of a location may depend on traffic density, the presence or not of traffic jams, work in progress, etc. that are all time dependent and usually hard to predict. Thus several approaches have been proposed to study parametric graphs [1] or dynamic and temporal graphs [6], [14].

Fully dynamic algorithms, for example, are applied to problems that can be solved in polynomial time. They start with a computed optimal solution, and then try to maintain them when changes occur in the problem. They often propose sophisticated data structures to reach this goal [8], [11].

When the delay between a change and the moment a new solution is needed is very small, or when the problem itself is NP-hard, faster algorithms are needed. These reoptimizing algorithms usually start from an initial solution that is not optimal but is expected to be of good quality, if possible. As soon as a change is detected, they compute a new solution, trying to do it faster that classical algorithms. Or they compute a new solution as fast as the classical algorithms but this resulting solution is better than the ones found by classical algorithms. These algorithms include meta-heuristics such as ants colony algorithms or swarm algorithms [3].

Another approach used is probabilistic. Probabilities are associated to some variables in the graph, such as the value of a weight, or the presence of a vertex or of a constraint, for example. The algorithms used in these problems usually compute a solution then do some robustness analysis in the probability space [9]. Or they do a quick re-optimization of the solution once the parameters of the problem are perfectly known [4], [10].

More specifically, on the computation of MST on Dynamic Graphs, several solutions are proposed: using partition and topology trees [15,16], using sparsification [17], using randomized algorithms [18], using logarithmic decomposition [19]. Almost all the proposed solutions follow the same general idea:

- an initial MST is computed for the initial state of the graph,
- an additional structure is added,
- this additional structure is used to re-compute as efficiently as possible the new MST as soon as a change is detected in the dynamic graph,
- this structure is usually itself updated too.

These solutions are developed for the most general case of fully dynamic graphs: anything can change, at any time.

In this paper, we study weakly dynamic undirected graphs, that can be used to represent some logistic networks.

For example, one might wish to "minimize the maximum" elevation change in a truck delivery problem: suppose that you are designing a path through a mountainous region. Since costs associated with traction power, and the wear and tear associated with braking, increase quickly with absolute elevation change, you may wish to choose an path that minimizes the maximum elevation change, and then add an additional cost for the duration of the trip.

This can be solved as a MST problem. However, an edge can see its weight changing during the trip (due to a traffic jam, an accident ...). In that case, a predefined MST is not minimal anymore. It seems interesting to anticipate these weight changes during the trip because the uncertainties of the network are known most of the time. This paper proposes to study that problem.

The network exists in a mostly stable environment, except for a few edges known to be non-stable. That is, the weight of each of these non-stable edges may change at any time (due to bascule or lift bridge, elevator, traffic congestion...).

## II. PROBLEM DESCRIPTION

Definition 1: A Weakly Dynamic Graph [7],[12] is a graph with valuated edges or arcs, in which there is one unstable valuated edge (in an undirected graph) or valuated arc (in a directed graph) between two known nodes $v1$ and $v2$ of the graph. That edge or arc has an unknown positive value $x$ that may change at any time. All other edges are stable and their values never change.

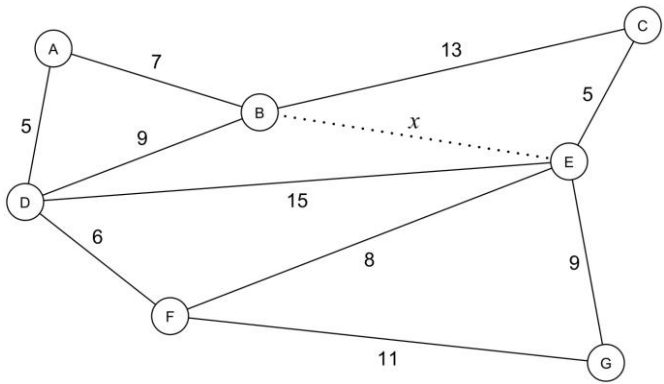

**Fig. 1** Example of a Weakly Dynamic Graphs The dashed line represents the unstable edge.

The general MST problem starts with a more general problem. Let $G = (V, E)$ be a simple undirected graph, with V being the set of vertices, and E being the set of edges of G.

Definition 2: an Edge-Constrained Spanning Tree *ECST (V, E, $E^+$, $E^-$)* of an undirected graph $G = (V, E)$ is a spanning tree of *G* with $E^+$ and $E^-$ being two disjoints subsets of *E*, and such that all edges in $E^+$ are in this spanning tree and no edge of $E^-$ is in this spanning tree.

In the following, we will call $E^+$ the set of mandatory edges, and $E^-$ the set of forbidden edges. We can notice that:

- this definition of mandatory is different from the definition of mandatory for an edge in MST in other papers, where an edge is said to be mandatory if all spanning trees must include it, else this spanning tree will not be minimal,
- the term "constrained" applied to a spanning tree has also the different meaning in other papers that some global criterion must be satisfied by the tree, such as a maximal weight, diameter or degree.

Of course, an edge-constrained spanning tree as we defined it may not always exist. For example, the subset $E^+$ of mandatory edges may already include a cycle, or the subset $E^-$ of forbidden edges may be such that the graph without its edges is not a connected graph anymore. Without loss of generality, we will suppose in the rest of this paper that one edge-constrained spanning tree actually exists.

An edge-weighted edge-constrained minimal spanning tree problem $P = (V, E, w, E^+, E^-)$, is a problem in which:

- *V* is a set of vertices,
- *E* is a set of edges $(i, j) | i \epsilon V, j \epsilon V$,
- *w* is a weight function with $w : E \to R$,
- $E^+$ is a subset of E of mandatory edges that must belong to the tree, $E^-$ is a subset of *E* of forbidden edges that are not allowed to belong to the tree.

Definition 3: a minimum (resp. maximum) spanning tree *T* in an edge weighted edge-constrained minimum (resp. maximum) spanning tree problem $P = (V, E, w, E^+, E^-)$ is an edge-constrained spanning tree of $G = (V, E)$ that verifies *ECST(V, E, $E^+$, $E^-$)* and such that no other spanning tree that verifies *ECST(V, E, $E^+$, $E^-$)* has a lower (resp. higher) weight.

## III. SOLUTION

### A. Preliminary results

- A solution of the classical minimum (resp. maximum) spanning tree problem is then a solution of the edge weighted edge-constrained minimum (resp. maximum) spanning tree problem $P = (V, E, w, \phi, \phi)$.
- Computing a minimum or maximum edge-weighted edge-constrained spanning tree without the forbidden edges of $E^-$ is trivial. Just build a subgraph of $G = (V, E)$ without these edges and apply an algorithm such a Prim algorithm [20], or Kruskal algorithm [21], (or any other).
- Computing a minimum or maximum edge-weighted edge-constrained spanning tree in a problem *P* with mandatory edges is more difficult.

First we suppose that $E^+$ has only one edge *(i, j)*. A modified Prim algorithm may then be used. In this algorithm, instead of starting from a random vertex, and without any initial edge in the tree, we may start with the initial set of vertices *{i, j}* and with the initial edge *(i, j)* in the tree and apply Prim algorithm from there.

Theorem 1: in an edge-weighted edge-constrained spanning tree problem $P = (V, E, w, E^+, E^-)$, if $E^+$ has only one edge, then the modified algorithm of Prim that starts from edge *(i, j)* of $E^+$ computes a minimum edge-constrained spanning tree.

Now we suppose that $E^+$ has two edges or more. The modified Prim algorithm cannot be used. We propose the following modified Kruskal algorithm. Instead of starting from an empty subset of *G* that will slowly be grown into a spanning tree, we may start with the initial subset $E^+$ and apply Kruskal algorithm from there.

Theorem 2: In an edge-weighted edge-constrained spanning tree problem $P = (V, E, w, E^+, E^-)$, if $E^+$ has one edge or more, then the modified Kruskal algorithm that starts with all edges *(i, j)* of $E^+$ computes a minimum edge-constrained spanning tree.

*B. The Proposed Algorithm*

We can now use the above result on a Weakly Dynamic Graph with one non-stable edge. Our proposed algorithm works in three steps:

1. First solve the minimum edge-constrained spanning tree problem $P = (V, E, w, \emptyset, \{(i, j)\})$, i.e. the problem without the non-stable edge. Its value is the sum of its edges $d_s$ and is a constant. This stable Minimum Spanning Tree is called $MST_s$. For example, by applying this on the graph of Fig. 1, we have:

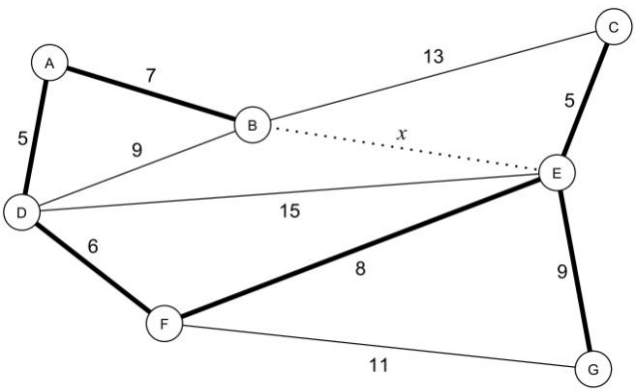

**Fig. 2** *Compute* of $MST_s$ without the non-stable edge x, The $MST_s$ value is the sum of its edges is a constant $d_s = 40$

2. Next solve the minimum edge-constrained spanning tree problem $P = (V, E, w, \{(i, j)\}, \emptyset)$, i.e. the problem with the mandatory non-stable edge. This variable Minimum Spanning Tree is called $MST_s$. Its value is the sum of its edges $d_v$ which is the sum of its stable edges, plus the non-stable value x, that may change at any time. By applying this on the graph of Fig. 1, we have:

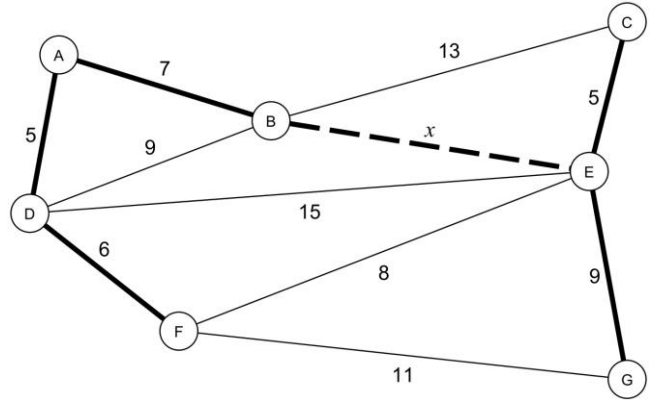

**Fig. 3** *Compute* of $MST_v$ with the non-stable edge x, The $MST_v$ value $d_v$ depends on $x$ : $d_v = 32 + x$

3. By comparing the values of $d_s$ and $d_v$, we can deduce the critical value *cv(x)* of *x*.

Definition 4: the critical value *cv(x)* of *x* is the value of *x* such that the minimum spanning tree will be either the first tree built (if $x > cv(x)$), or the second one (if $x < cv(x)$).

In our example, we have: $d_s = 40$, $d_v = 32 + x$, thus the critical value $cv(x) = 8$.

If $x < cv(x)$ then the best *MST* is the $MST_v$ including the non-stable edge x. If $x >= cv(x)$ then the best *MST* is the $MST_s$ (without the non-stable edge *x*).

In the example of Fig. 1, one can see that if *x* is greater than 8, then the $MST_s$ computed in step 1 (and presented Fig. 2) is a correct MST. And if *x* is lower than 8, then the $MST_v$ computed in step 2 (and presented Fig. 3) is a correct MST.

These two MST are then alternative MST depending on the current value of *x*.

*C. Algorithm complexity*

About the complexity of our algorithm, the complexity of Prim's algorithm = $O(|V|^2)$ with a choice of the appropriate data structure, the complexity can be reduced to $O(|E| + |V| \log |V|)$ [5], in a similar way, the complexity of Kruskal's algorithm can be reduced to $O(E \log E)$. Thus, our algorithm complexity is $O(n^2)$, with $n = |V|$.

IV. REMARKS AND CONCLUSION

Because all MSTs are pre-computed, the two trees built, and the critical value *cv(x)*, may be stored somewhere and no re-computation is needed each time the non-stable edge changes

- either the new value of *x* is on the same side of the critical value *cv(x)* than the old value of *x*, and no change is needed,

- or the new value of $x$ is on the other side of the critical value $cv(x)$ than the old value of $x$, and the other precomputed spanning tree will be used instead of the current one.

Thus, the response time is the best possible.

In our example, the MSTs of Fig. 2 and Fig. 3 will be stored, and the critical value of 8 for $x$ will be used to instantly determine which one will be used, without any recomputation. These results may be extended to a small number of non-stable edges, but the combinatorial nature of this problem makes it impractical even for a medium number of non-stable edges.

However, if we suppose that only one change is possible at any time on one of several non-stable edges, and that the delay between two changes is long enough, then after each change we can solve $m$ ($m$ being the number of non-stable edges) separate one non-stable edge problems (with the corresponding trees and critical values for each non-stable edge) and again have the best possible response time when the next change occurs in one of the $m$ non-stable edge.

We are currently trying to improve these pre-computations.

ACKNOWLEDGMENT

This project is co-financed by the European union with the European regional development fund and by the Normandy regional council. Projet : CLASSE "Corridors Logistiques : Applications à la vallée de la Seine et Son Environnement", France.